# Automated Enumeration of Reconfigurable Architectures for Thermal Management Systems in Battery Electric Vehicles

Reihaneh Jahedan[1], Satya Peddada [2], Mark Jennings[3], Sunil Katragadda[3], James Allison[2]*,

Nenad Miljkovic[1,4,5,6,7,8]*

[1]Mechanical Science and Engineering, University of Illinois at Urbana-Champaign, Urbana, IL 61801, USA

[2] Industrial and Enterprise Systems Engineering, University of Illinois at Urbana-Champaign, Urbana, IL 61801, USA

[3] Ford Motor Company, Dearborn, MI 48126, USA

[4]Materials Research Laboratory, University of Illinois at Urbana-Champaign, Urbana, IL 61801, USA

[5]Electrical and Computer Engineering, University of Illinois at Urbana-Champaign, Urbana, IL 61801, USA

[6]International Institute for Carbon Neutral Energy Research (WPI-I2CNER), Kyushu University, Fukuoka, 819-0395, Japan

[7]Institute for Sustainability, Energy and Environment (iSEE), University of Illinois, Urbana, Illinois, USA

[8]Air Conditioning and Refrigeration Center (ACRC), University of Illinois, Urbana, Illinois, USA

*Correspondence: jtalliso@illinois.edu, nmiljkov@illinois.edu

**Abstract**

As the automotive industry moves towards vehicle electrification, designing and optimizing thermal management systems (TMSs) for Battery Electric Vehicles (BEVs) has become a critical focus in recent years. The dependence of battery performance on operating temperature, the lack of waste combustion heat, and the significant effect of TMS energy consumption on driving range make the design of BEV TMSs highly complicated compared to conventional vehicles. Although prior research has focused on optimizing the configuration of thermal systems for varying ambient conditions, a holistic approach to studying the full potential of reconfigurable TMS architectures has not yet been fully explored. The complex design landscape of multi-mode reconfigurable systems is difficult to navigate. Relying solely on expert intuition and creativity to identify new architectures both restricts progress and leaves significant performance improvements unrealized. In this study, using graph modelling of TMS architectures, we propose a systematic method to automatically enumerate and simulate reconfigurable architectures for a TMS, given the desired operating modes, along with a framework to conduct transient performance analysis and optimization-based trade-off studies among system performance, energy consumption, and complexity. We explored more than 150 operating mode sequences, retaining 39 unique architectures for further evaluation. MATLAB Simscape models of these architectures were automatically created and their performance evaluated. The multi-objective optimization results provide decision support for selecting the best architecture based on user priorities.

## Keywords

# Nomenclature

**Abbreviations**

| | |
|---|---|
| BEV | Battery Electrical Vehicle |
| TMS | Thermal Management System |
| BTMS | Battery Thermal Management System |
| HP | Heat Pump |
| AC | Air Conditioning |
| NLP | Non-Linear Programming |
| DT | Drive Train |
| LCC | Liquide Cooled Condenser |
| US06 | High acceleration aggressive driving schedule |
| GB | Gear Box |
| MC | Motor Controller |
| COP | Coefficient Of Performance |

**Symbols**

| | |
|---|---|
| $T$ | Temperature (°C) |
| $T_{DT}^{max}$ | Maximum temperature of the drivetrain components including front and rear motor, gear box and motor controller. (°C) |
| $T_{cabin}^{initial}$ | Temperature of the cabin at the beginning of the simulation. (°C) |
| $T_{battery}^{initial}$ | Temperature of the battery at the beginning of the simulation. (°C) |
| $T_{DT}^{initial}$ | Temperature of the drivetrain at the beginning of the simulation. (°C) |
| $T_{cabin}^{final}$ | Temperature of the cabin at the end of the simulation. (°C) |
| $T_{battery}^{final}$ | Temperature of the battery at the end of the simulation. (°C) |
| $T_{cabin}^{comfort}$ | The cabin temperature defined to calculate heating time of cabin. (°C) |
| $G_{\text{architecture}}$ | The graph representing the TMS architecture |
| $V'$ | Set of nodes in $G_{architecture}$ representing thermal components or junctions. |
| $E'$ | Set of edges in $G_{architecture}$ representing coolant lines. |
| $E'_{valve}$ | Subset of edges in $G_{architecture}$ representing coolant lines with inline vales. |
| $G_{\text{constraint}}^{i}$ | The directed graph representing thermal constraints of operating mode $i$ |
| $V$ | Set of nodes in $G_{constraint}$ representing main thermal components used in the system. |
| $E^i$ | Set of edges in $G_{constraint}$ representing necessary heat transfers in operating mode $i$ |
| $(A, B)$ | An edge connecting node A to node B |
| $C_{A,B}$ | A cycle passing through node A and node B |
| $p$ | The node representing pump in $G_{architecture}$ |
| $path_G\,((A,B),(C,D))$ | A path in graph $G$, connecting edges $(A,B)$ and $(C,D)$ |
| P | Mechanical power (W) |
| t | Time (seconds) |
| $t_{Cabin}^{heat}$ | Time required to heat the cabin to the $T_{cabin}^{comfort}$. |

**Subscripts**

| | |
|---|---|
| a | active (using heat pump) |

p

passive (using radiator or waste heat recovery)

# 1. INTRODUCTION

In recent years, battery electric vehicles (BEVs) have gained popularity due to a combination of environmental concerns, rising fuel costs, and governmental support. Despite advances in battery technology and charging infrastructure that have made BEVs more convenient for everyday use, limitations such as the short lifespan of the battery and its sensitivity to temperature variations, coupled with the restricted driving range of BEVs, continue to hinder widespread adoption. Studies have shown that a well-designed thermal management system (TMS) is crucial to maximizing battery performance and lifespan by maintaining optimal temperature levels. Furthermore, an efficient TMS can extend the driving range of a BEV, since TMS energy consumption constitutes a significant portion of the total BEV energy use [1]. Given the critical role of TMS in advancing BEV adoption, researchers and industry professionals have focused on improving the design and efficiency of BEV TMS[2–7].

Several comprehensive reviews [7–9] provide detailed information on advancements in BEV TMS technologies. Research efforts in this area can be categorized into component and system level optimizations. Component level studies have investigated specific technologies such as battery thermal management systems (BTMS) [8–14], refrigerants [15,16], heat pumps (HP) [17], and air conditioning (AC) systems [18,19]. For example, Kim *et al*. [9] reviewed and categorized BTMS technologies for lithium-ion batteries, summarizing their advantages and limitations. Based on these insights, they proposed a combined BTMS solution that integrates multiple technologies to complement the shortcomings of individual approaches for high-energy-density lithium-ion batteries. In another study, Wang et al. [13] employed a multidisciplinary design optimization framework of variable fidelity to refine the BTMS design parameters. Similarly, refrigerant-based systems, such as AC and HPs, have been optimized. Wang *et al*. [20] tested an integrated AC/HP system with two refrigerants, analyzing the effects of refrigerant type, compressor speed, and ambient temperature on heat capacity and energy efficiency. On the other hand, system-level strategies focus on the holistic integration of components and optimized design and control mechanisms. Some system-level studies [21–23] proposed optimized controllers for EV AC systems to enhance energy efficiency and cabin thermal comfort. Min *et al*[24] explored integrated TMS control approaches [24] where a fuzzy control strategy improved battery life without significantly compromising cabin and battery temperatures. During the last decade, researchers and industry professionals have introduced several novel approaches to holistic TMS configurations [25–30], emphasizing the importance of multimodal BEV TMS systems that accommodate diverse weather and operating conditions[26,28,29].

Among system-level approaches, Singh *et al*. [26] conducted a systematic evaluation of the design trade-offs of BEV TMS designs using a virtual testbed. They proposed multiple sets of operating modes to optimize system efficiency under varying ambient and operating conditions, employing waste heat recovery. Although these operating modes and control strategies have been optimized in multiple studies, the design of reconfigurable architectures to enable desired configurations has mainly relied on engineers' intuition and experience. This reliance underscores a critical gap in the literature: the lack of systematic methods to explore, evaluate, and optimize reconfigurable TMS architectures. A decision support tool is developed in this study to address this gap. The first step used to explore the design space in the proposed approach is an enumerative method that can efficiently generate multiple design candidates. These design candidates can then be evaluated, and in the final step, users can select the design that best meets their specific needs. This structured approach ensures a systematic exploration of the design space, moving beyond reliance on intuition.

The enhancement of Non-Linear Programming (NLP) algorithms enabled automation and optimization in a variety of design steps, including the selection of design variables [31], the tuning of the control parameters [32], and the evaluation of the system[33,34]. Even though most of the design automation problems are considered continuous optimization problems, system architecture design automation is mostly considered a discrete optimization problem and relies on automated enumerative algorithms. The system architecture design automation has been investigated in various domains, ranging from molecule synthesis [35] to on-chip interconnects [36], robotics [37], powertrain design [38], and TMSs [39–43]. Several studies have modeled architectures as graphs and employed graph-theoretic concepts to enable enumeration and analysis. For example, Herber *et al.* [44] represented the architecture design space as graph sets and proposed an automated architecture enumeration method utilizing a perfect matching technique to enumerate all possible architectures for a moderate-scale system design problem. This method was extended to explore automated architecture design and evaluation for aircraft TMS [43,45]. Although their method successfully enumerated and modeled TMS architectures for aircraft, it did not focus on generating reconfigurable architectures capable of supporting multiple operating modes. Instead, like Singh *et al.* [26], their configurations were optimized for specific operating points, determining how components should be connected under fixed conditions. Graph-based modeling approaches proposed in architecture enumeration studies form a foundation for the systematic exploration of reconfigurable BEV TMS architectures. However, they require adaptation to address the dynamic demands, and reconfiguration needs specific to BEV TMS.

Multi-modal reconfigurable architectures have been studied extensively in other domains, such as manufacturing [46–49], robotics [50–53], powertrain design [54,55], and multi-purpose sensor placements [56,57]. These studies emphasize the importance of designing systems capable of adapting to changing operational conditions, often employing principles like modularity and dynamic topology reconfiguration. In the context of robotics, for example, Sihite *et al.* [51] designed a robot capable of multiple modes of mobility. Utilizing its multi-functional components, this robot can fly, roll, crawl, crouch, balance, tumble, scout, and loco-manipulate. Similarly, reconfigurable powertrain systems have been designed to enable multiple optimized modes, allowing vehicles to adapt to varying operating weights and payloads [54]. In addition to proposing novel multi-modal designs, formalizing and capturing the overall concept and design process of multi-modal systems have also been investigated [50,54,58]. For instance, Tan *et al.* [50] proposed an eight-step design process for developing multi-modal bio-inspired robots. However, since the specific dynamics and needs of each system highly affect the design criteria, the proposed methodologies are tailored to domain-specific challenges and are not directly transferable to BEV TMS design. This underscores the need to develop reconfigurable design methodologies specific to the unique requirements of BEV TMS, including managing thermal loads, optimizing energy consumption, and adapting to diverse operating conditions.

While system architecture enumerative design and reconfigurable system design have been extensively studied in other domains, their application in BEV TMS design remains largely unexplored, and a holistic approach aimed at minimizing architectural complexity has been absent. To address this gap, this article introduces a novel approach for automated reconfigurable TMS architecture design. The proposed methodology is tailored to accommodate the unique challenges of BEV TMS, enabling multiple operating modes while minimizing architectural complexity.

The design of BEV TMSs presents unique challenges due to the need for high efficiency, adaptability to varying operating conditions, and minimal energy consumption. While significant progress has been made in component-level and system-level optimization, the exploration and evaluation of reconfigurable architectures for BEV TMSs remain largely unstructured and reliant on expert experience and intuition. This article proposes a systematic framework to facilitate the exploration and evaluation of reconfigurable BEV TMS architectures, supporting early-stage design decisions. Given a set of desired operating modes tailored for varying weather and operating conditions, multiple reconfigurable architectures are automatically designed and modeled in MATLAB Simscape. The generated models are simulated across diverse test case scenarios, and multiple metrics are utilized to evaluate and compare the enumerated architectures using multi-objective optimization techniques.

The primary objective of this work is to develop a systematic and automated methodology for the design and performance evaluation of reconfigurable TMS architectures tailored for BEVs. Addressing the limitations of intuition-driven design practices, this study introduces a graph-based framework capable of enumerating and simulating multimodal TMS configurations to support diverse operational scenarios. The proposed approach enables comprehensive trade-off analysis among system performance, energy consumption, and architectural complexity. The major contributions of this work are as follows:

- Development of an automated graph-based enumeration method for generating reconfigurable BTMS architectures based on a set of desired operating modes.
- Integration of constraint-based logic to ensure that the enumerated architectures satisfy thermal flow requirements, coolant isolation, and flow restrictions for each operating mode.
- Automatic generation of dynamic simulation models for all enumerated architectures, incorporating preprocessing techniques such as valve placement, pump configuration, and node simplification.
- Implementation of simulation-based performance assessment, including both fixed and dynamic test cases, to evaluate thermal regulation, energy consumption, and cabin heating responsiveness.

- Formulation of a multi-objective evaluation framework, enabling trade-off analysis between architectural complexity and performance using Pareto front visualizations.
- Enumeration and evaluation of 39 unique reconfigurable architectures, offering insights into the structural-performance relationships and supporting early-stage design decision-making.

This framework not only supports the scalable exploration of BEV TMS configurations but also lays the groundwork for future studies in automated design of reconfigurable thermal systems across various engineering domains.

The remainder of this manuscript is organized as follows. Section 2 of this article discusses the proposed enumerative architecture design method. Section 3 elaborates on the modeling of these architectures in Simscape. Results are presented and discussed in detail in Section 4. The final section of the article, Section 5, concludes with a summary of contributions and potential future research topics.

The outcomes of this research have the potential to significantly impact the BEV industry by providing a new approach to TMS design. By exploring reconfigurable architectures, this study aims to contribute to the development of more efficient, scalable, and adaptable TMS designs for BEVs. This, in turn, could reduce the weight and initial cost of the thermal system, increase battery lifespan, extend driving range, and reduce operational costs, thereby accelerating the broader adoption of BEVs. Moreover, the research methodology and findings could serve as a foundation for future studies in TMS design and related fields.

## 2. DESIGN METHODOLOGY

### 2.1 Operating Modes

In an EV TMS, the operating modes dictate the essential heat transfer processes among system components, designed to meet the thermal demands of varying operational conditions. A control system determines the active operating mode based on critical parameters including ambient temperature, battery temperature, drive train (DT) temperature, and cabin temperature. The DT includes an integrated electric motor, motor controller and gearbox assembly. In this study, the TMS configuration and a total of 13 operating modes, suggested by Singh *et al* [26], were adopted. These operating modes were carefully selected to encompass a wide range of real-world BEV usage scenarios, including six winter modes, six summer modes, and a specific charging mode. Winter and summer modes account for seasonal extremes and high-load aggressive driving conditions, where a drive cycle defines the vehicle speed profile that directly influences heat generation in the battery and DT. In contrast, the charging scenario (mode 7) simulates conditions where the cabin remains unoccupied, eliminating the need for cabin temperature maintenance and modeling of passengers' heat generation. Additionally, no heat is produced in the DT during charging, and battery heat generation is modeled specifically according to charging conditions. This comprehensive set ensures that the enumerated architectures effectively support the thermal demands encountered throughout the operational envelope of the vehicle. A summary of these operating modes is presented in Table 1, and two of them are depicted in Figure 1. The main thermal components used in the suggested configuration include a HP that delivers low temperature refrigerant to a chiller and high temperature refrigerant to a liquid cooled condenser (LCC), a cabin heat exchanger, a battery heat exchanger, cooling passages integral to the DT assembly, and two radiators.

## 2.2 Graph Representation

The architecture of a BEV TMS includes details of the overall coolant lines connecting thermal components, their intersections, and the valves and pumps integrated within the system. In this study, graphs were used to model both the TMS architectures and the operating modes. In the architecture graphs, as shown in Figure 2, nodes represent thermal components and junctions, while edges signify coolant lines. To simplify the enumeration process, pumps are added to the architecture in a post processing step. Moreover, only inline on-off switch valves were considered. Coolant lines with these valves were highlighted in the architecture graph using colored edges.

**Table 1:** The ambient and operating conditions of the ideal operating modes and heat transfer requirements A: active (using HP) P: passive (using radiator or waste heat recovery)

| Season | Operating Mode | Ambient [ºC] | Cabin [ºC] | Battery [ºC] | DT [ºC] | Heat Transfer |
|---|---|---|---|---|---|---|
| Winter | Mode 1 | $T < 20$ | $T < 20$ Heating$_A$ | $T < 15$ Heating$_A$ | $T_{DT}^{max} < 25$ Neutral | LCC → Cabin; LCC → Battery; Radiator1 → Chiller |
| | Mode 2 | $T < 20$ | $T < 20$ Heating$_A$ | $T < 15$ Heating$_P$ | $T_{DT}^{max} > 25$ Cooling | LCC → Cabin; DT → Battery; Radiator1 → Chiller |
| | Mode 3 | $T < 20$ | $T < 20$ Heating$_A$ | $25<T<40$ Cooling$_P$ | $T_{DT}^{max} > 25$ Cooling | LCC → Cabin; DT → Radiator2; Battery → Radiator2; Radiator1 → Chiller |
| | Mode 4 | $T < 20$ | $T < 20$ Heating$_A$ | $25<T<40$ Cooling$_P$ | $T_{DT}^{max} < 25$ Neutral | LCC → Cabin; Battery → Radiator2; Radiator1 → Chiller |
| | Mode 5 | $T < 20$ | $T < 20$ Heating$_A$ | $15<T<25$ Neutral | $T_{DT}^{max} < 25$ Neutral | LCC → Cabin; Radiator1 → Chiller |
| | Mode 6 | $T < 20$ | $T < 20$ Heating$_A$ | $15<T<25$ Neutral | $T_{DT}^{max} > 25$ Cooling | LCC → Cabin; DT → Chiller |
| Charging Mode | Mode 7 | $T < 20$ | Empty Neutral | $T < 15$ Heating$_A$ | Off Neutral | LCC → Battery; Radiator1 → Chiller |
| Summer | Mode 8 | $T > 20$ | $T > 20$ Cooling$_A$ | $25<T<40$ Cooling$_P$ | $T_{DT}^{max} > 35$ Cooling | LCC → Radiator 1; Cabin → Chiller; Battery → Radiator2; DT → Radiator2 |
| | Mode 9 | $T > 20$ | $T > 20$ Cooling$_A$ | $T > 40$ Cooling$_A$ | $T_{DT}^{max} > 35$ Cooling | LCC → Radiator 1; Cabin → Chiller; Battery → Chiller; DT → Radiator2 |
| | Mode 10 | $T > 20$ | $T > 20$ Cooling$_A$ | $25<T<40$ Cooling$_P$ | $T_{DT}^{max} < 35$ Neutral | LCC → Radiator 1; Cabin → Chiller; Battery → Radiator2 |
| | Mode 11 | $T > 20$ | $T > 20$ Cooling$_A$ | $T > 40$ Cooling$_A$ | $T_{DT}^{max} < 35$ Neutral | LCC → Radiator 1; Cabin → Chiller; Battery → Chiller |
| | Mode 12 | $T > 20$ | $T > 20$ Cooling$_A$ | $T < 25$ Neutral | $T_{DT}^{max} < 35$ Neutral | LCC → Radiator 1; Cabin → Chiller |
| | Mode 13 | $T > 20$ | $T > 20$ Cooling | $T < 25$ Neutral | $T_{DT}^{max} > 35$ Cooling | LCC → Radiator 1; Cabin → Chiller; DT → Radiator2 |

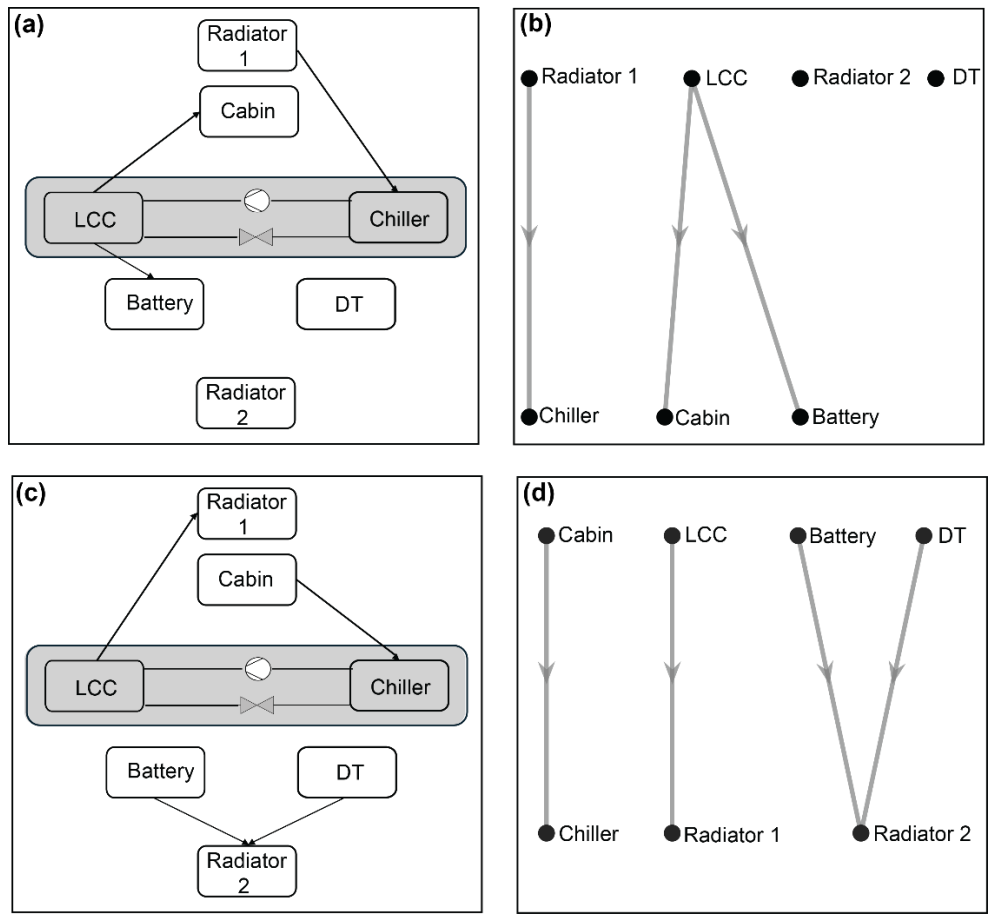

**Figure 1**: Two of the selected operating modes. **(a)** Mode 1 is a winter operating mode in which the cabin and battery both need heating, and the DT does not need any cooling or heating. **(b)** Graph representation of mode 1. **(c)** Mode 8 is a summer mode in which the cabin needs active cooling by HP, and the battery and DT can use passive cooling through the radiator. **(d)** Graph representation of mode 8. Arrows indicate the direction of heat transfer between components. Schematics not to scale.

After the enumeration, these valves can get replaced with multi-way valves commonly used in practical applications. For each operating mode, the necessary heat transfers were depicted using a directional graph, referred to as a constraint graph. As shown in Figure 1, nodes in these graphs represent thermal components, while directional edges illustrate the heat transfer between them. For example, if heat transfer is required from the Cabin Heat Exchanger to the Chiller in a specific operating mode, a directional edge connects the corresponding nodes in that mode's constraint graph.

Mathematically, the architecture graph can be represented as $G_{\text{architecture}} = (V', E')$, where V' is the set of thermal components, junctions, and pumps, and E' is the set of edges representing coolant lines. The subset of edges representing coolant lines with inline valves is denoted as $E'_{\text{valve}} \subseteq E'$, and these are shown as red edges in the architecture graph. The constraint graph for operating mode $i$ is represented as $G^i_{\text{constraint}} = (V, E^i)$ where $V$ is the set of thermal components and $E^i$ is the set of directed edges representing the required heat transfers in the operating mode $i$.

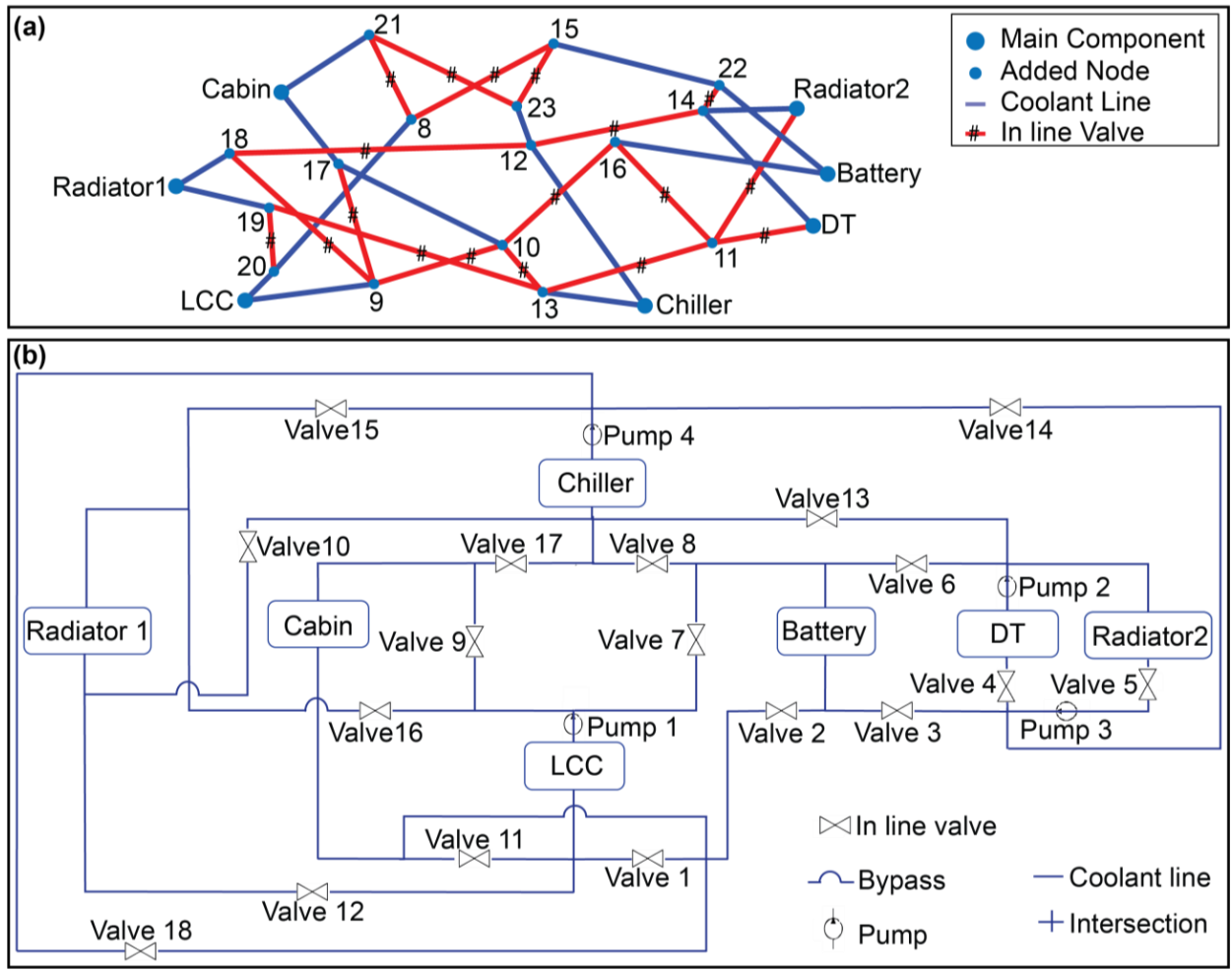

**Figure 2:** BEV TMS architecture graph. **(a)** Code-generated architecture graph. **(b)** Same architecture with simpler visualization.

## 2.3 Problem Formulation

A TMS architecture can enable an operating mode if each heat transfer requirement in the mode is supported by a corresponding coolant loop within the architecture. Given a necessary heat transfer between two components, A and B, this coolant loop must pass through components A, B, and at least an active pump p. The coolant within this loop must remain isolated from other coolants in the system and must not flow to any component that is not connected to A or B in the operating mode graph. In other words, the conditions for a TMS architecture to enable the operating mode $i$ are as follows:

1. Existence of a Cycle:
$$\forall (A,B) \in E^i \Rightarrow \exists \{C_{A,B} \subseteq G_{\text{architecture}} | A, B, p \in C_{A,B}\}. \quad (1)$$

2. Coolant Isolation:
$$\{(A,B),(C,D) \in E^i | \nexists \, path_{G^i_{\text{constraint}}}((A,B),(C,D))\} \Rightarrow C_{A,B} \cap C_{C,D} = \emptyset. \quad (2)$$

3. Flow Restriction: The coolant in cycle $C_{A,B}$ must not flow through any components that are not directly connected to $A$ or $B$ in $G_{\text{constraint}}$. Therefore, if a component $k$ is in $C_{A,B}$, then $k$ must be either $A$, $B$, or directly connected to $A$ or $B$ in $G_{\text{constraint}}$. Mathematically:

$$\forall k \in C_{A,B} \Rightarrow k \in \{\{A,B\} \cup \{N|(A,N) \in E^i \text{ or } (A,B) \in E^i\}. \quad (3)$$

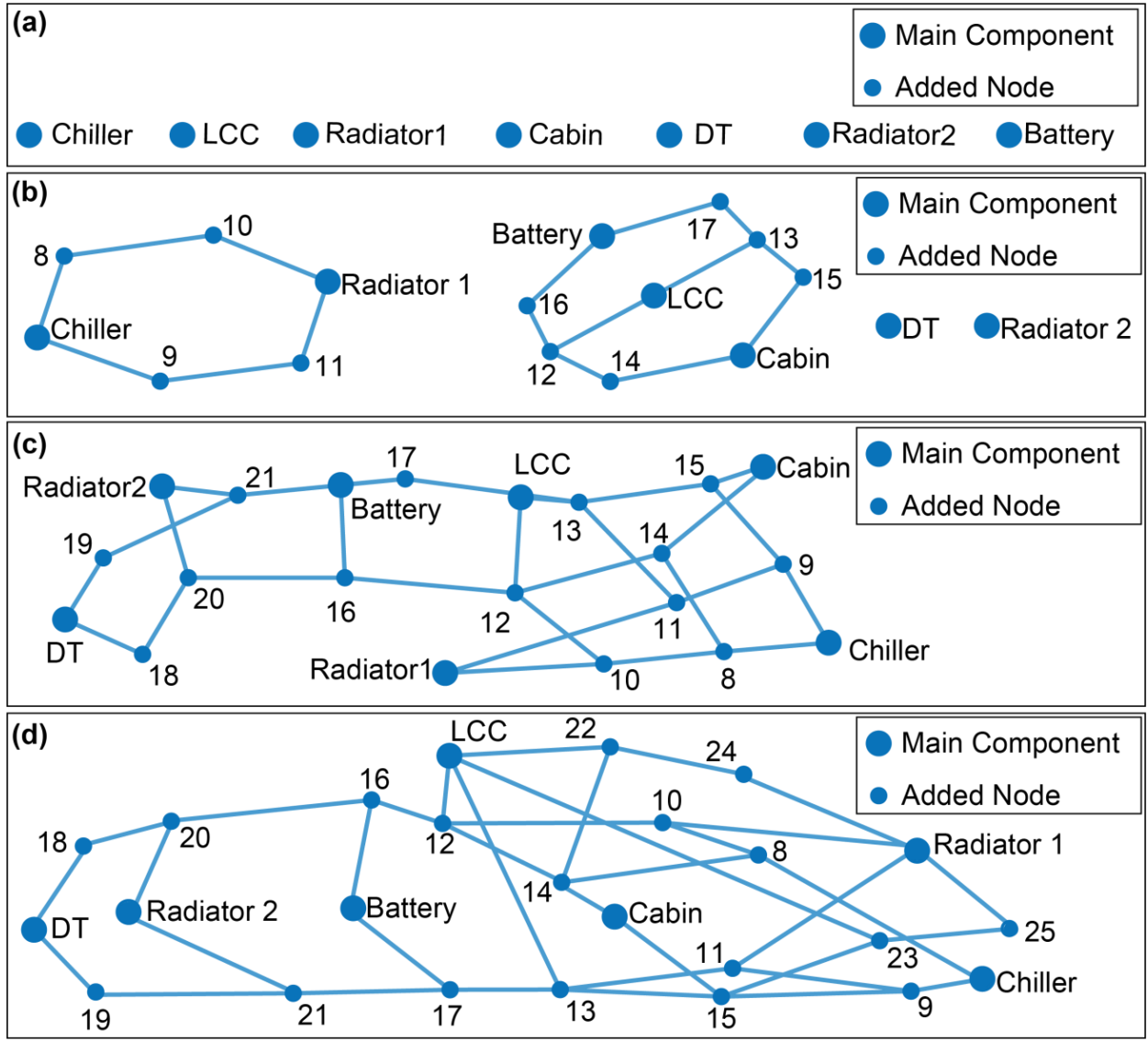

**Figure 3:** Architecture Enumeration Process. **(a)** Architecture Graph after the initialization process. **(b)** Architecture graph, after adding the necessary coolant lines for mode 1. **(c)** Architecture graph, after adding mode 8 on top of mode 1. **(d)** Architecture graph including necessary coolant lines for all of the modes, enumerated following this random sequence of modes: 1,8,11,2,5,9,10,3,4,7, 12,6,13

## 2.4 Automated Layered Architecture Enumeration Method

The proposed layered architecture enumeration process includes five steps that are described in the following sub-sections.

### 2.4.1   Initialization:

The enumeration method constructs the final architecture by sequentially enabling operating modes and adding them to the architecture graph, layer by layer. Since the sequence in which modes are added affects the final architecture, a random sequence was generated during the initialization process. To generate multiple unique architectures for the TMS, a set of different sequences was applied. Initially, the architecture graph contains only seven nodes, representing the main thermal components of the TMS. Figure 3.a, shows this initial architecture graph.

### 2.4.2 Adding coolant lines:

Given the sequence generated during the initialization step, the constraint graph of the operating modes, and the initial architecture graph, the process of adding coolant lines to the architecture graph begins by sequentially incorporating operating modes according to the predefined order. At each step, the incorporated mode is designated as the active mode, and its corresponding constraint graph is called the active constraint graph. To incorporate the active operating mode into the architecture graph, the architecture graph is scanned for the required cycle based on Equation (1) for each edge in the active constraint graph. If no such cycle exists, nodes and edges are added to the architecture graph to form the necessary cycle. Additionally, to satisfy Equation (2), the cycles for separate pairs of components in the active constraint graph must not share any edges, ensuring distinct coolant paths for each set of connections. This process is depicted in Figure 3. Once valid cycles have been identified for all edges in the active constraint graph, the procedure is repeated for the subsequent active mode, following the initial sequence. This method continues until the required edges for all operating modes are successfully added to the architecture graph. Different architectures are generated when coolant lines are added to the architecture graph following different sequences. In summary:

For each random sequence:

1. Start from the initial graph (the initial set of seven components).
2. Consider operating modes one-by-one, in the order specified by the sequence as the active operating mode.
    a. For each active operating mode, identify the connections required between components based on the active constraint graph
    b. For each edge in the active constraint graph (necessary heat transfer in the active operating mode):
        i. Check if the current graph (architecture) already has a suitable cycle (loop) to allow coolant flow.
        ii. If no suitable loop exists, add new nodes and edges (coolant lines) to create one.
        iii. Ensure that the cycles created for the unconnected edge pairs in the active constraint graph do not overlap or share nodes or edges in the architecture graph. Each required heat transfer must have a distinct coolant path.
3. Continue this process until coolant lines have been successfully added for every operating mode in the sequence.
4. Repeat the process following other sequences to generate new architectures.

2.4.3 Unique Architecture Selection:

Given the set of architecture graphs generated in the previous step, some of the architectures enumerated from different sequences might be like each other. Unique architectures were determined by removing the isomorphic graphs from the enumerated set of architecture graphs. In this process to distinguish the main thermal components, nodes representing them were color-coded. In the first trial of this method, 39 unique architectures were enumerated after trying 150 different sequences.

2.4.4 Adding Valves:

After identifying unique architecture graphs, valves are added to each architecture graph to enable the coolant isolation and flow restriction constraints as represented in Equation (*2*) and Equation (3) respectively. The following steps are repeated for each unique enumerated architecture graph. For each operating mode, the nodes and edges in the architecture graph that are used by each coolant are preserved from the previous step, where coolant lines were added to the architecture. Some nodes might remain unused by any coolant flows. Multiple options are then evaluated for assigning these unused nodes to specific coolant flows. Subsequently, different cut sets, defined as a set of edges in a graph that, if removed, there is no path from any node in one group to any node in the other, are determined. Cut sets can divide the architecture graph into disconnected sub-graphs, based on the active constraints, and therefore must be determined. All potential unions of these cut sets for all operating modes are identified, and the smallest union is selected as the valve set of the architecture. The cut set used for each operating mode in the selected valve set is also saved to determine the valve status (on or off) in each operating mode. The selected valve set is highlighted in the graph using colored edges, indicating the placement of inline on-off switch valves in the TMS architecture. In summary:

For each unique architecture:

1. For each operating mode:

    a. Coolant paths are identified in the graph.

    b. Unused nodes are determined and assigned to different coolant flow paths.

    c. Cut sets that can disconnect different coolant flow paths in the architecture graph are determined.

2. Consider every possible combination of valve placements across all operating modes.

3. Select the smallest set of valves that fulfills all coolant isolation and flow-control requirements for all operating modes.

4. Highlight the valve placement using colored edges in the architecture graph.

## 3. MODELING

MATLAB Simscape was employed to model and evaluate the performance of enumerated TMS architectures. Simscape, a MATLAB toolbox designed for simulating physical systems, offers a component-based approach with predefined blocks for thermal components like heat exchangers, valves, and pumps, making it well-suited for TMS simulations that require precision and efficiency. This section details the steps involved in the automatic model creation process, including preprocessing and adjustments to enable accurate simulation in Simscape.

### 3.1 Graph-Based Pre-processing:

The adjacency matrices representing the enumerated architectures require several pre-processing steps to facilitate their direct translation into Simscape models. These steps are necessary to ensure the model reflects the physical properties of the system accurately. The process is pictured in Figure 4 and the associated steps are elaborated below:

1. Conversion to Directed Graphs: In the enumeration phase, an undirected graph was used to represent the architecture, which allowed bidirectional flow paths for flexibility. This approach was suitable for generating various architectural layouts. However, for accurate model creation in Simscape, a directed graph is required to correctly assign inlet and outlet nodes, especially because certain components, like pumps, are unidirectional. Pumps in the TMS are designed to only allow coolant flow in a single direction, making it essential to set up clear flow paths from input to output in the simulation model. While some components, such as coolant valves and some primary system components (e.g., Battery, and DT), are physically capable of handling bidirectional flow, the Simscape blocks representing these components require designated inlet and outlet ports for correct simulation. By converting the undirected graph from the enumeration phase into a directed graph for the simulation phase, we ensure that each component is correctly assigned a functional directionality. This step guarantees that the pumps are configured to pump coolant in the appropriate direction and that the flow orientation is accurately represented across all components.

2. Elimination of Redundant Nodes: During the enumeration process, auxiliary nodes were sometimes introduced to facilitate the formation of coolant flow cycles within the graph. However, some of these nodes, identifiable by having only two connections (degree of two), do not correspond to any actual junction or component in the physical TMS. These nodes are thus considered redundant and are removed from the graph to ensure that the model accurately reflects the physical layout of the system.

   When removing a redundant node, its neighboring nodes (those directly connected to it) are reconnected to maintain continuous flow paths within the architecture. If any valve is associated with the edges connected to the redundant node, the valve and its status (open or closed) must be preserved in the reconnected path.

3. Pump Placement: To maintain coolant circulation, each coolant loop requires at least one active pump. For simplification, pumps are restricted to placement at either the inlet or outlet of a thermal component. Initial tests across various operating modes identified that four pumps were sufficient to support all required coolant flow paths for the modeled architectures. Pumps were strategically placed at the following locations: the inlet of the chiller, the outlet of the LCC the inlet of the second radiator, and the outlet of the DT. In this preprocessing step, these pumps are represented as additional nodes within the directed architecture graph.

4. Incorporation of Valves as Nodes: Inline valves were initially represented as colored edges within the graph, with their positions determined by cut sets. However, since modified Simscape blocks are used to model the valves for simulation, each valve must be represented as a node in the graph rather than an edge. In this step, each colored edge representing a valve is replaced by a node, and the new valve node is connected to the endpoints of the removed edge. This adjustment ensures that flow paths remain continuous and that valve status information (open or closed) is preserved, enabling accurate simulation across various operating modes.

5. Node Labeling: To ensure consistency between the architecture graph and the Simscape model, each node is assigned a label that reflects its function within the TMS matches the corresponding Simscape block. In this step, labels assigned to key components (such as main thermal units, valves, and pumps) are verified to ensure correctness, as these components were labeled earlier in the process.

(a) Thermal Components: Main thermal components (e.g., radiators, chillers, battery thermal systems) are labeled with their specific names to directly correspond with the designated Simscape blocks for these units.

(b) Valves and Pumps: Valves and pumps, which were also labeled in previous steps, are checked to ensure that each one is correctly numbered and matches its intended placement within the TMS.

(c) Intersection Nodes: T-junctions and Cross Junctions are the standard form of intersections used in this study. Nodes representing intersections are assigned labels based on their degree which shows the number of branches they connect. Intersection nodes with a degree of three are labeled as T-Junctions, while those with a degree of four are labeled as Cross-Junctions. The intersection nodes that have a higher degree are substituted by a set of connected T Junctions and Cross Junctions that would correctly mimic the branch connection. For instance, an intersection node with a degree of five will be replaced by a T Junction and a Cross Junction that are connected to each other.

By verifying and standardizing labels across all nodes, this step ensures that each component and connection in the Simscape model accurately reflects the architecture, aiding in both the simulation setup and interpretation of results.

## 3.2 Assessment

Before proceeding to modeling and simulation, it was crucial to verify that the enumeration method and preprocessing steps succeeded, ensuring the modified architectures could meet the requirements of each operating mode defined in Equation (1), Equation (2), and Equation (3). To accomplish this, we developed an algorithm that verified each architecture against the operational requirements. For each architecture and mode, the algorithm removed the off valves from the graph and checked that:

1. Each edge in the constraint graph of the operating mode was supported by a cycle in the architecture graph that included the main thermal components and an active pump (Equation (1))
2. No cycle with an active pump includes components that should remain unconnected in the given mode (Equation (2), Equation (3)).

The assessment step ensured the integrity of the modified architectures before their use in subsequent modeling and simulation stages. A sample result of this algorithm that visualizes the coolant path in mode 1 in one of the enumerated architectures is shown in Figure 5.

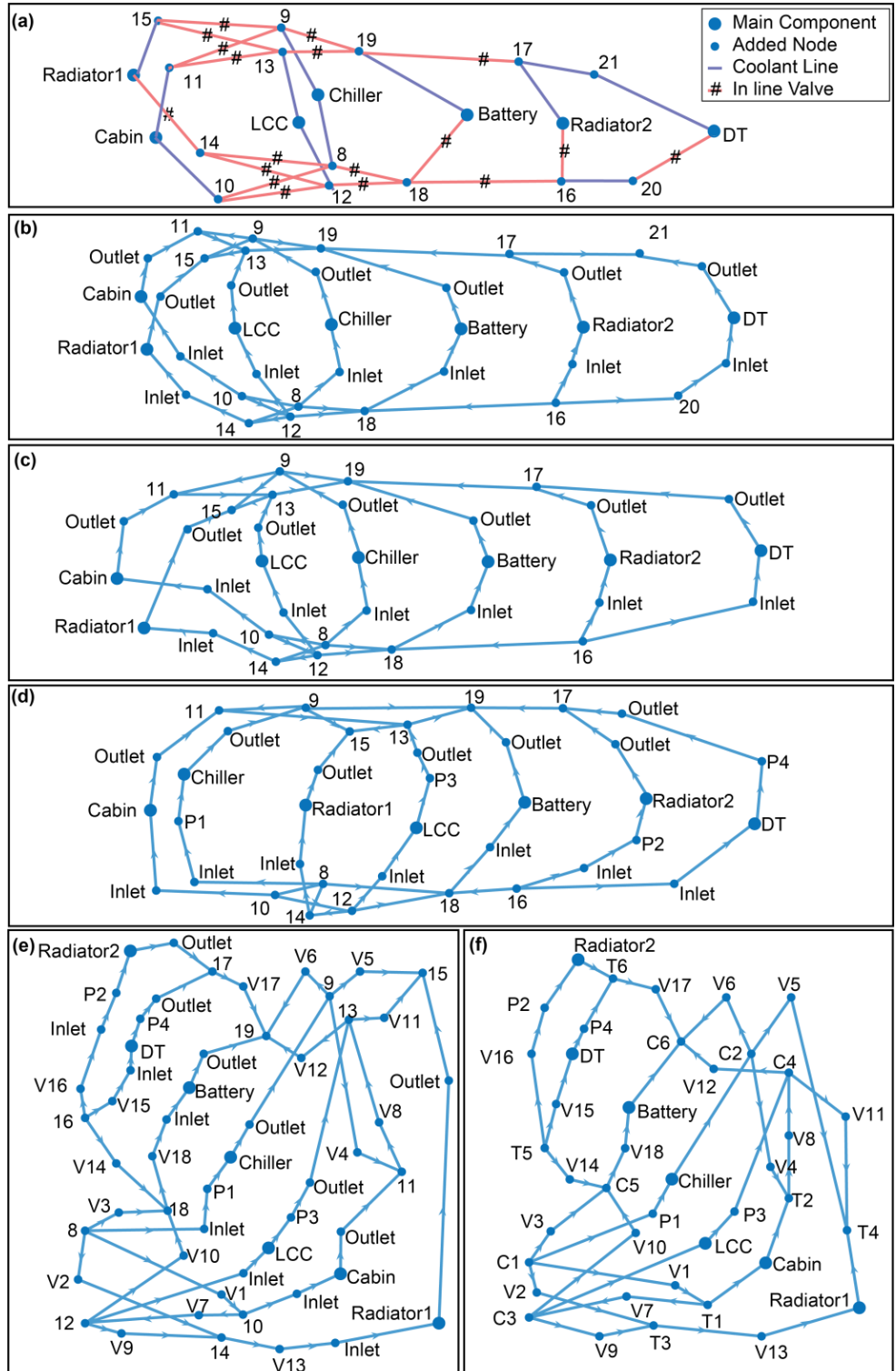

**Figure 4:** Graph-Based Pre-Processing. **(a)** Code enumerated graph: red edges show the edges with an inline valve on them. **(b)** Architecture directional graph: Inlet and outlet of main components are highlighted using inlet and outlet nodes. **(c)** Redundant Nodes (Node 20 and 21) Removed from the directional graph **(d)** Pumps added to the architecture directional graph. **(e)**

Valve nodes added to the architecture graph. **(f)** Final architecture directional graph ready for automated modeling. The inlet and outlet nodes are removed, and all nodes are labeled properly.

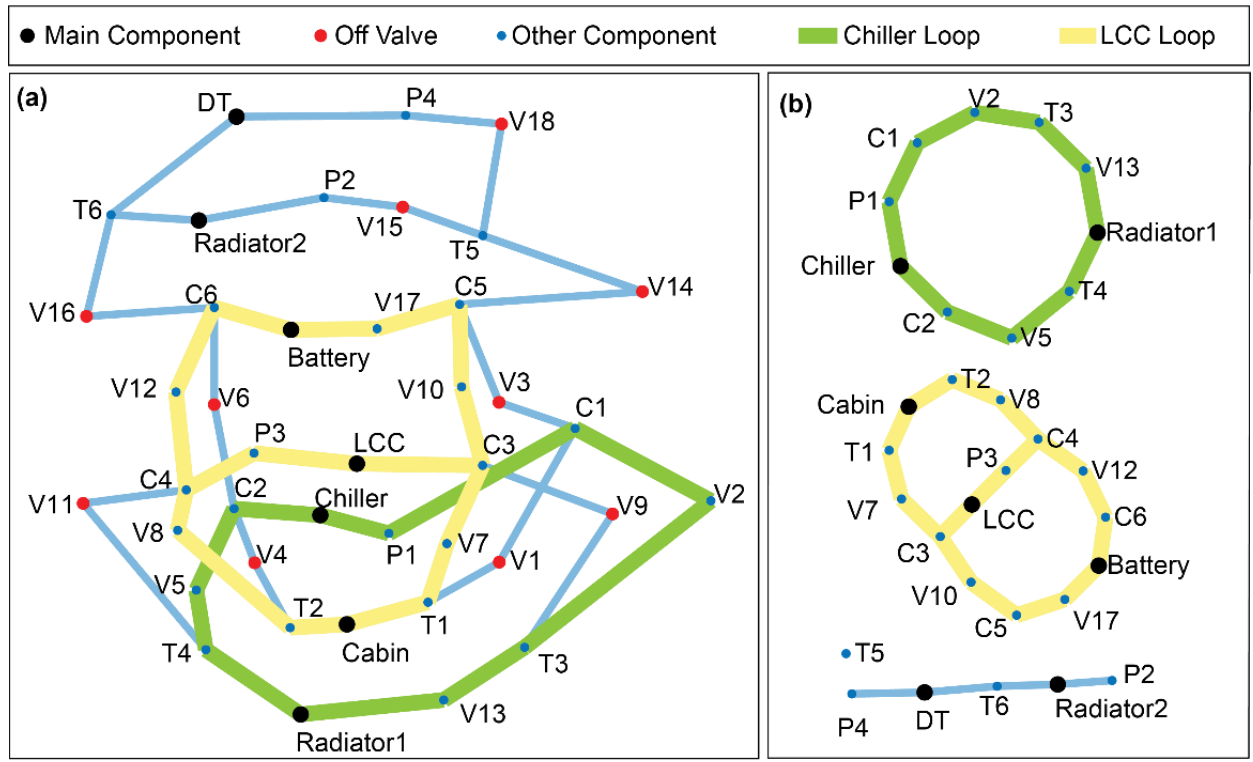

**Figure 5:** The Assessment result for architecture 1 in mode 1. Red nodes in the original graph show the off valves. **(a)** Original TMS Architecture. **(b)** Modified TMS Architecture (off valves removed)

### 3.3 Model Creation

A custom Simscape library was developed to provide the essential component models necessary for simulating each TMS architecture. This library includes customized thermal models of key components, adapted from Singh *et al.* [26], as well as modified versions of certain predefined Simscape blocks and a control section for managing system operation modes. Singh *et al.* [26] developed and validated Simscape models for the primary thermal components of a BEV TMS (including the HP, radiators, cabin thermal system, battery thermal system, and DT thermal system) considering heat loads across a range of drive cycles. These core models are supplemented by modified standard Simscape blocks representing additional components, such as valves, junctions, pumps, and tanks, which have been adapted to ensure compatibility with the control section.

To efficiently model each enumerated architecture each architecture is generated using a script that assembles the necessary subsystems from the Simscape library and interconnects them based on the preprocessed adjacency matrix. This automated scripting ensures consistent and accurate model creation without requiring manual adjustments.

The control section, also configured through scripting, enables dynamic switching between operating modes by controlling the states of valves and pumps. Stored valve and pump states from the enumeration process are used to set up the control section for each model, allowing each architecture to flexibly adapt to different operational conditions. When a valve is closed, it blocks flow in the associated line, while an inactive pump functions as a simple bypass, maintaining the continuity of the coolant loop. These modifications to the standard blocks make them fully compatible with the automated control logic.

Each model is simulated across a range of ambient temperatures and drive cycles to evaluate the thermal performance of the enumerated architectures. This script-driven approach to model creation, configuration, and control ensures efficient scalability and reproducibility across various TMS configurations.

## 3.4 Simulation and Test Cases

To evaluate the performance of the generated TMS architectures, two sets of tests were conducted: Fixed Operating Mode Tests and Dynamic Operating Mode Tests. These tests assess the ability of each system to manage cabin and battery temperatures of a four-wheel drive full size pickup truck under various operational conditions and evaluate their energy efficiency. The simulated vehicle configuration and related component parameters are the same as that used by Singh *et al*. [26].

1. Fixed Operating Mode Tests:

    Each operating mode was tested individually, with the active mode held constant throughout the simulation. The operating modes were designed to respond to specific weather and vehicle conditions, as outlined in Table 1. In this test, initial temperatures for the cabin, battery, and DT were set based on these conditions. These initial temperatures are shown in Table 2. The simulations modeled 10 minutes of vehicle operation using the US06 drive cycle. Since operating mode number 7 is designed for thermal management during battery charging, it was not simulated in this test case. Key performance metrics are calculated using Equation ((4)), Equation ((5)) and Equation ((6)). Where $T_{cabin}^{final}$, $T_{battery}^{final}$ respectively show the temperature of the cabin and the battery at the end of the simulation. Similarly $T_{cabin}^{initial}$, $T_{battery}^{initial}$ respectively show the initial temperature of the cabin and the battery. The temperature changes in the cabin ($\Delta T_{cabin}$) and the battery ($\Delta T_{battery}$) are among indicators of architecture performance.

$$\Delta T_{cabin} = T_{cabin}^{final} - T_{cabin}^{initial}. \tag{4}$$

$$\Delta T_{battery} = T_{battery}^{final} - T_{battery}^{initial}. \tag{5}$$

    Energy consumption is another performance metric calculated using Equation (6). The consumed energy is the summation of energy consumption of the compressor and pumps. The $P_{compressor}$ in this equation refers to the mechanical power consumption of the compressor, $P_{pump}^{i}$ shows the mechanical power consumption of pump $i$, and $t$ represents time in seconds.

Minimized energy consumption is desirable since it saves energy and extends the driving range of the vehicle.

$$\text{Total Energy} = \int_{t=0}^{t=t_{\text{end}}} P_{\text{compressor}}(t)dt + \sum_{i=1}^{4} \int_{t=0}^{t=t_{end}} P_{\text{pump}}^{i}(t)dt \,. \tag{6}$$

Lower energy consumption and a greater change in cabin temperature indicate better performance across all operating modes. The desired battery temperature change varies depending on the specific purpose of each operating mode. Some modes are designed to heat the battery and raise its temperature, while others aim to maintain or cool it and prevent overheating during operation. The results section provides a detailed assessment of the architecture's performance under these varying objectives.

2. Dynamic Operating Mode Test:

This test simulated a more realistic scenario. Here, the initial temperatures of the cabin, battery, and DT matched the ambient temperature of 273 K (or 0°C). Vehicle operation was modeled over a 30-minute period encompassing three repetitions of the US06 drive cycle. A cabin temperature set point of 293 K (or 20°C) was targeted. Performance was measured by how fast each architecture can heat the cabin and the total energy consumption (Total Energy). Since in the simulation, the cabin temperature asymptotically approaches the desired set point and might not reach it numerically in the simulation, the comfort temperature is calculated using Equation (7). The comfort temperature ($T_{cabin}^{comfort}$) is defined as 90% of the temperature difference between the temperature set point ($T_{cabin}^{set\ point}$) and the initial temperature ($T_{cabin}^{initial}$) added to the initial temperature .

$$T_{cabin}^{comfort} = 0.9(T_{cabin}^{setpoint} - T_{cabin}^{initial}) + T_{cabin}^{initial}. \tag{7}$$

The time required to reach the cabin comfort temperature ($t_{cabin}^{heat}$), is measured using Equation(8).

$$t_{Cabin}^{heat} = \min(t) \text{ such that } T_{cabin}(t) > T_{cabin}^{comfort}. \tag{8}$$

$t_{cabin}^{heat}$ is used as a performance metric. Lower $t_{cabin}^{heat}$ is considered to be superior performance since it means the desired cabin temperature is achieved faster. This increases the cabin comfort and passenger satisfaction.

The results of these simulations, detailed in the following section, provide insights into the performance of each TMS architecture, highlighting critical trade-offs in energy efficiency and thermal management.

**Table 2:** Ambient temperature and component (cabin, battery and DT) initial Temperature for the static fixed operating mode tests. The modes are defined in detail in the work of Singh. *et al*[26].

| Operating Mode | Ambient Temperature [$^{o}$C] | Cabin Initial Temperature, $T_{\text{cabin}}^{\text{initial}}$ [$^{o}$C] | Battery Initial Temperature, $T_{\text{battery}}^{\text{initial}}$ [$^{o}$C] | DT Initial Temperature, $T_{\text{DT}}^{\text{initial}}$ [$^{o}$C] |
|---|---|---|---|---|
| Mode 1 | 0 | 0 | 0 | 0 |
| Mode 2 | 0 | 0 | 0 | 26 |
| Mode 3 | 0 | 0 | 26 | 26 |
| Mode 4 | 0 | 0 | 26 | 0 |
| Mode 5 | 0 | 0 | 16 | 0 |
| Mode 6 | 0 | 0 | 16 | 26 |
| Mode 8 | 35 | 35 | 35 | 51 |
| Mode 9 | 35 | 35 | 41 | 51 |
| Mode 10 | 35 | 35 | 35 | 35 |
| Mode 11 | 35 | 35 | 41 | 35 |
| Mode 12 | 35 | 35 | 24 | 35 |
| Mode 10 | 35 | 35 | 24 | 51 |

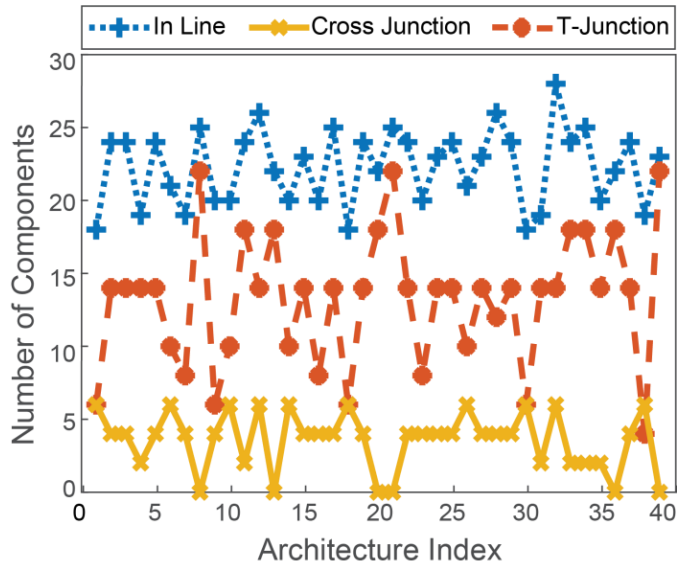

**Figure 6:** The number of in-line valves, T-junctions, and cross junctions used by each of the enumerated architectures.

## 4. RESULTS AND DISCUSSION

This section presents and discusses the performance of the generated TMS architectures based on the evaluation metrics outlined in the previous section. The results are structured to highlight key aspects of the architectures, including their complexity, energy efficiency, and thermal regulation performance under fixed and dynamic operating conditions. After testing 150 sequences of the 13 operating modes adopted from Singh *et al* [26], 39 unique architectures were enumerated and modeled in this study. To facilitate a comprehensive analysis, the findings are organized as follows:

### 4.1 Architectural Complexity

The enumerated architectures incorporate varying numbers of valves, T-junctions, and cross-junctions. These components contribute to system complexity and have several practical implications. Increased complexity results in higher initial costs and added weight, potentially impacting the overall vehicle performance. Moreover, a greater number of components can reduce the system's robustness by increasing the likelihood of leakage or failure. The added complexity also poses challenges for control implementation, making it more expensive and difficult to design reliable control algorithms. Figure 6 illustrates the number of in-line valves and junctions used in each architecture, providing a visual comparison of their complexity.

### 4.2 Fixed Operating Mode Test Performance

As described in the previous section, the fixed operating mode test evaluates the performance of each architecture for pre-selected fixed operating modes. This analysis confirms that all architectures can function across the designated modes and compares their relative performance. This approach is particularly beneficial when certain operating modes have greater significance in the design objectives. For instance, Singh *et al* [26] showed that certain operating modes occur more frequently. Therefore, assigning higher weight to these modes is important during the design of the TMS. Additionally, in operating modes typically active during the vehicle's initial phase, the rate of cabin temperature change is a more critical metric compared to modes where the primary goal is maintaining cabin temperature. Another situation where assigning different weights to operating modes is beneficial occurs when the TMS is designed for vehicles intended for markets with specific climate conditions.

The capability of each architecture to regulate cabin temperature is presented in Figure 7, which depicts the absolute values of cabin temperature change ($\Delta T_{\text{cabin}}$) after ten minutes of operation. Superior cabin temperature regulation performance corresponds to larger absolute values of $\Delta T_{\text{cabin}}$. The results indicate greater temperature changes when the HP exclusively manages cabin temperature, as opposed to operating modes where it concurrently regulates both cabin and battery temperatures.

Interestingly, architectures that exhibit superior cabin temperature regulation in winter do not necessarily perform best in summer conditions. For example, Figure 7b indicates that in most summer operating modes, including mode 8, architecture 36 outperforms architecture 39. Conversely, Figure 7a demonstrates that this relationship does not hold for several winter modes, such as mode 5. Since the HP exclusively manages the cabin temperature in both mode 5 (winter) and mode 8 (summer), these two modes were selected to compare the performance of architectures 36 and 39 in different seasons. Figure 8 illustrates the cabin temperature profiles for architectures 36 and 39 in modes 5 and 8. The results show that architecture 36 achieves superior cooling performance in summer by reducing cabin temperature faster, whereas architecture 39 heats the cabin more quickly in winter. This finding highlights the importance of carefully prioritizing specific operating modes during design, as performance trade-offs across diverse conditions significantly affect the overall suitability of a given architecture.

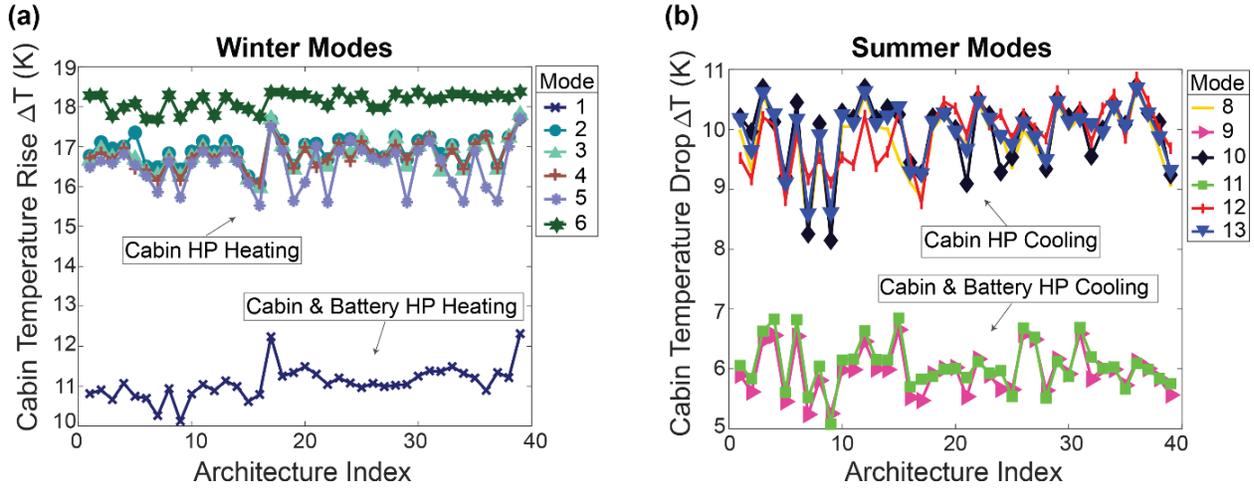

**Figure 7:** Cabin temperature change by the end of each fixed operating mode test. Different lines show different operating modes. **(a)** Cabin temperature rise in winter. The cabin temperature increased more when exclusively heated by the HP. **(b)** Cabin temperature drop in summer. The cabin temperature decreased more when exclusively cooled by the HP.

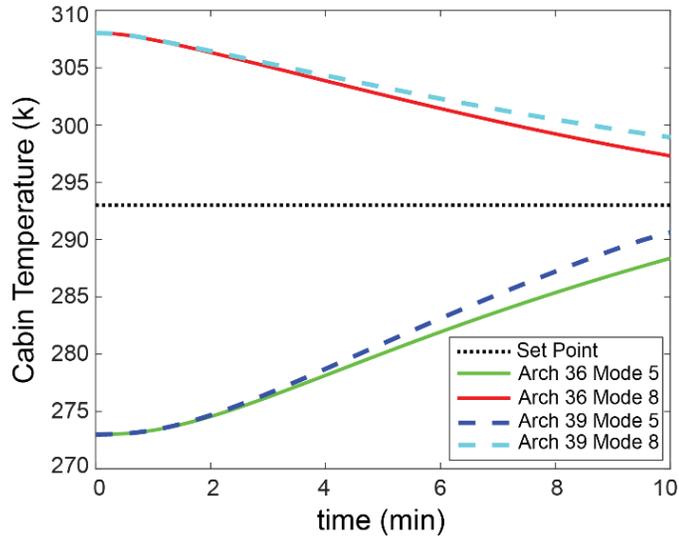

**Figure 8:** Cabin Temperature Profile Comparison in architecture 36 and architecture 9 for a winter mode (mode 5) and a summer mode (mode 8).

Battery temperature rise across individual operating modes is shown in Figure 9. As mentioned, regarding battery temperature regulation, superior performance depends on the design objective of each operating mode. In modes where the battery requires heating (modes 1 and 2), a larger difference between the initial and final battery temperature indicates better performance. In modes where the battery requires cooling (modes 3, 4, 8, 9, 10, and 11), the TMS must offset the heat generated by the battery during operation. Therefore, better performance corresponds to a smaller difference between the battery's initial and final temperature. In certain operating modes (modes 5,6,12,13), it is assumed that the battery remains within the desired temperature range. Therefore, thermal regulation is not required. In these cases, the preferred outcome is for the battery temperature to remain close to its initial value.

Data indicates that for most architectures, the highest battery temperature rise occurs in mode 2, where waste heat from the DT is used for heating. However, the effectiveness of mode 2 varies across architectures. This mode is unsuitable for immediate use after vehicle startup in winter, since it requires the DT to be sufficiently warm. Architectures that achieve a higher temperature rise in this mode are best suited for scenarios where the DT generates heat more rapidly, allowing frequent activation of mode 2. The second-highest temperature rise is observed in mode 1, where the battery is heated by the HP. As mentioned previously, in both modes, the battery starts below its optimal operating range, making rapid heating a priority.

In the remaining operating modes, the architecture exhibiting the lowest battery temperature rise is considered the best performing. The small temperature rise observed in these modes ensures the battery remains within the proper range. As expected, active cooling (used in modes 9 and 11) and passive cooling during winter conditions (modes 3 and 4) result in better performance across all architectures, compared to passive cooling during summer (modes 8 and 10) or neutral modes.

The differences in battery temperature rise within each operating mode are generally insignificant (less than 1ºC), making it difficult to draw meaningful performance distinctions between the architectures. The only operating modes in which architectures had different performance in terms of battery temperature rise is mode 2, in which some architectures such as architecture 33 had a better performance compared to some other architectures such as architecture 7.

Energy consumption varies significantly across different modes and architectures, as illustrated in Figure 10a. The results show that the minimum energy consumption occurs in winter mode 6, where waste heat from the DT is used as a heat source by the HP system. This efficiency of the mode underscores the importance of leveraging waste heat for energy conservation.

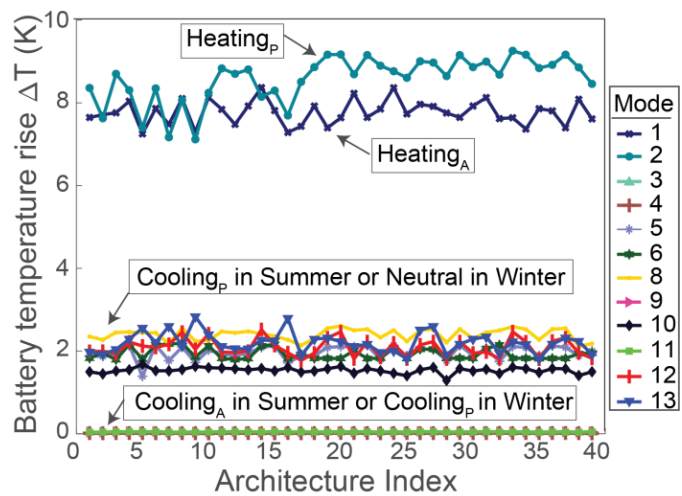

**Figure 9**: The difference between maximum temperature and starting temperature of the battery for the fixed operating mode tests. Different lines show different operating modes. A-active, P-passive. Passive heating refers to heating the battery using DT waste heat recovery. Passive cooling refers to using the radiator to cool the battery.

The heating and cooling capacities of the system vary considerably between modes, making it difficult to draw consistent conclusions about energy consumption differences for specific architectures across all modes. Within individual modes, meaningful comparisons can be made. For instance, in most winter modes (except mode 1), architectures 17 and 39 have the lowest energy consumption, while architectures 7 and 16 exhibit the highest. Interestingly, the variation in energy consumption across architectures is more pronounced in winter modes compared to summer modes. This suggests that the performance of certain architectures is more sensitive to operating conditions in winter, likely due to the increased heating demands and reliance on waste heat recovery circuits.

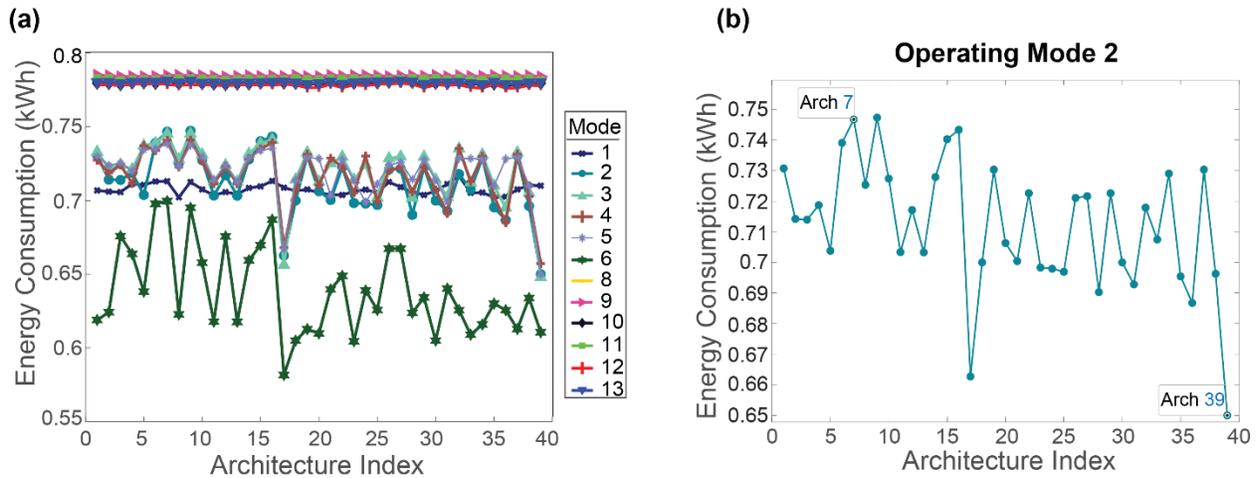

**Figure 10: (a)** Total energy consumption of each architecture for the fixed operating mode tests. Different lines show different operating modes. **(b)** Total energy consumption of each architecture in mode 2.

Analyzing energy consumption variations across different architectures and operating modes requires examining compressor power, pumping power, heat transfer rates in heat exchangers, and coolant paths. Operating mode 2 is selected as a representative example for presentation here. In this mode, the HP exclusively manages cabin temperature, simplifying the analysis of the compressor load, the primary energy consumer. Figure 10b provides a clear comparison of energy consumption across architectures in operating mode 2. Architecture 39 exhibits the lowest energy consumption, whereas architecture 7 shows the highest. For these two architectures, Table 3 summarizes total energy consumption, compressor energy usage, and details regarding the coolant paths, including the number of valves, T-junctions, and cross-junctions in the LCC and Chiller loops. Visual representations of the coolant paths for architectures 7 and 39 operating in mode 2 are provided respectively in Figure A. 1 and Figure A. 2 of Appendix A. Since most energy consumption occurs within the HP, the waste heat recovery coolant loop connecting the DT to the battery was excluded from this analysis, as it primarily affects pumping power, which is negligible in this context.

In the architectures modeled in Simscape, coolant passing through gate valves experiences a pressure drop; however, the impact of this pressure drop on overall energy consumption is negligible, as it primarily affects pumping power. In contrast, junctions with unused branches significantly influence energy efficiency. At these junctions, a small portion of coolant flows into unused branches, causing coolant loss and consequently reducing the overall loop efficiency due to wasted thermal energy. Specifically, in architectures 7 and 39 during operating mode 2, cross junctions had two unused branches, amplifying coolant waste. Since architecture 7 includes more junctions and unused branches in both chiller and LCC loops, it is expected that the HP will operate less efficiently in this architecture. Figure 11c compares the waste in LCC loop for architectures 7 and 39 throughout the simulation, calculated by subtracting the coolant mass flow rate in the cabin heat exchanger from that in the LCC. As shown in Figure 11b, architecture 7 exhibits a lower heat transfer rate in the cabin due to higher coolant waste. Conversely, architecture 39 achieves a higher heat transfer rate, leading to a more rapid cabin temperature increase. This reduces the difference between cabin temperature and the desired setpoint, subsequently decreasing the PI controller's input signal controlling the compressor. Ultimately, as observed in Figure 11a, the compressor turns off sooner in architecture 39, significantly reducing its overall energy consumption.

**Table 3:** Energy metrics and coolant path characteristics for architectures 7 and 39 under operating mode 2 conditions.

| Architecture | Energy Consumption | Lcc Loop | Chiller Loop |
|---|---|---|---|

|   |   | Compressor Energy Consumption | # Valve | # T Junction | # Cross Junction | # Valve | # T Junction | # Cross Junction |
|---|---|---|---|---|---|---|---|---|
| 7 | 0.76 | 0.73 | 2 | 2 | 2 | 2 | 4 | 2 |
| **39** | 0.65 | 0.63 | 3 | 4 | 0 | 2 | 4 | 0 |

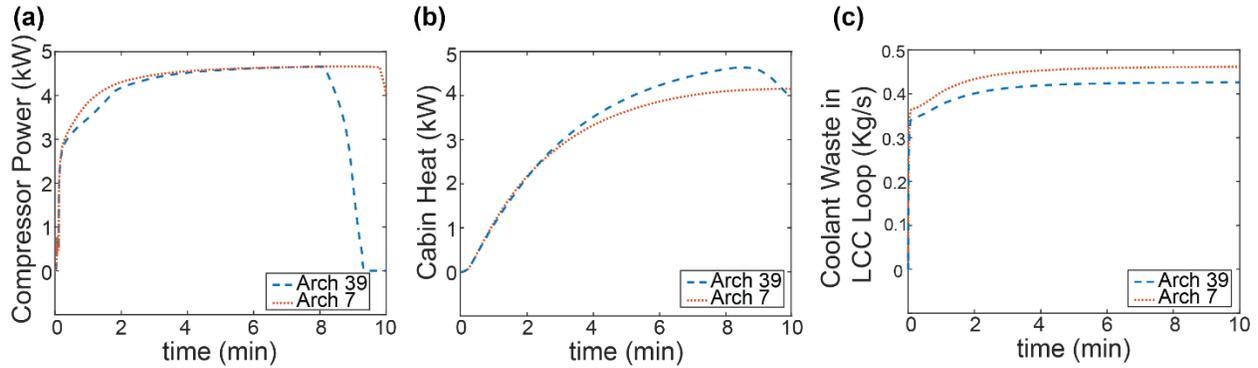

**Figure 11**: Performance comparison of architectures 7 and 39 in operating mode 2: **(a)** Compressor power consumption over the simulation period; **(b)** Heat transfer rate in the cabin heat exchanger; **(c)** Coolant waste in the LCC loop, calculated as the difference between coolant mass flow rates in the LCC heat exchanger and the cabin heat exchanger, representing flow diverted into unused branches at junctions.

Although mode analysis was not the primary objective of the fixed operating mode test, the observed cabin and battery temperature changes, along with energy consumption across different modes, align with the findings of Singh *et al*. [26], further validating the reliability of the developed models. While mode analysis demonstrates how a single architecture's performance varies across different modes, the primary focus of this study was to compare the overall performance of different architectures which will be discussed in dynamic operating mode test.

**4.3 Dynamic Operating Mode Test:**

This test evaluates the performance of the architectures during on road driving conditions. The battery, cabin, and DT temperature responses, as well as the sequence of activated operating modes, follow a similar trend across all architectures. Figure 12 shows the system response for architecture 1 as a sample. Operating mode 1 is activated at vehicle startup, during which the cabin and battery are heated by the HP. Once DT warms up, mode 2 is triggered, allowing the battery to be heated using DT waste heat. In this mode, the HP focuses solely on heating the cabin, resulting in an accelerated cabin temperature rise. When the battery reaches its optimal operating temperature and no longer requires heating, mode 6 is activated. In this mode, DT waste heat is delivered to the chiller to enhance the coefficient of performance (COP) of the HP.

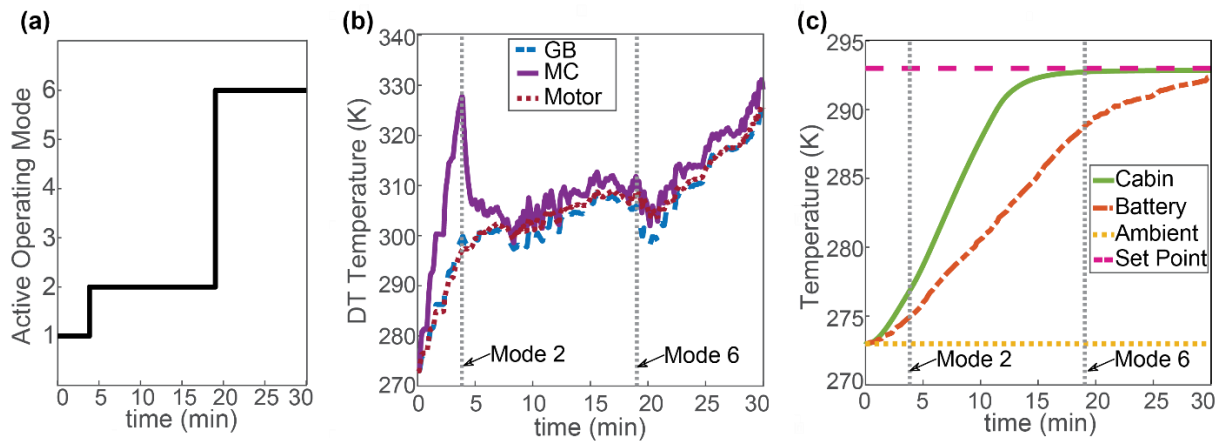

**Figure 12:** Active operating modes and component temperatures for architecture 1 during the dynamic operating mode test. Similar trends were observed across other architectures, with slight variations in mode switch timing. **(a)** Active mode during the simulation. **(b)** DT temperature profile. GB refers to gearbox and MC shows the motor controller temperature **(c)** Battery and cabin temperature profile.

Although this sequence is consistent across all architectures, significant differences are observed in the overall energy consumption calculated from Equation (6) and the time it takes to reach the cabin comfort defined by Equation (8). These variations, depicted in Figure 13, emphasize the influence of architectural design on dynamic performance metrics.

As shown, architectures 17 and 39 exhibited the best performance in both metrics, achieving faster cabin heating while maintaining lower energy consumption. Architectures 5, 11, 13, and 21 followed as the next set of low energy consumption designs. However, among them, only architecture 5 demonstrated noticeably better performance in terms of cabin heating rate, whereas architecture 21 performed the worst in this regard.

Given the challenge of evaluating multiple performance metrics alongside the structural complexity of each architecture, a multi-objective approach is essential for a comprehensive assessment. To better understand the trade-offs between performance and architectural simplicity, a Pareto front analysis is presented in the following section.

**4.4 Pareto Front Analysis:**

The fixed and dynamic tests effectively evaluated and compared the enumerated architectures across various performance criteria. However, supporting informed design decisions requires a holistic, multi-objective approach that carefully balances key trade-offs. Such analysis typically involves defining a robust utility function, informed by lifecycle assessments, cost analyses, and customer preference studies. Developing this comprehensive utility function for the architectures enumerated in the current study is beyond the scope of this article. Therefore, a simplified optimization analysis is presented in this section to demonstrate the proposed method's capability as an early-stage decision-support tool.

For clarity and ease of visualization, this optimization considers only two objectives, representing the complexity and performance of the architectures. Additionally, as noted previously, component sizing and capacity were fixed based on findings from previous studies [26]. Consequently, no inner-loop optimization was conducted to determine optimal component sizing for each architecture. Instead, this study exclusively focuses on identifying the most effective reconfigurable interconnections for the predefined operating modes.

The total number of valves and junctions was used as the complexity factor. Performance evaluation relied on the dynamic operating mode test case, as it better reflects actual on road system operation. Since all architectures achieved the cabin and battery temperature target set points, performance metrics were based on total energy consumption and cabin heating time. To enable better visualization, these metrics were normalized and summed to form a single performance objective function. Equation shows this calculation for architecture $j$. This approach can be adapted to accommodate different design priorities and considerations.

$$Performance\ Objective(j) = \frac{Total\ Energy(j)}{max\ (Total\ Energy)} + \frac{t_{cabin}^{heat}(j)}{max(t_{cabin}^{heat})}. \quad (9)$$

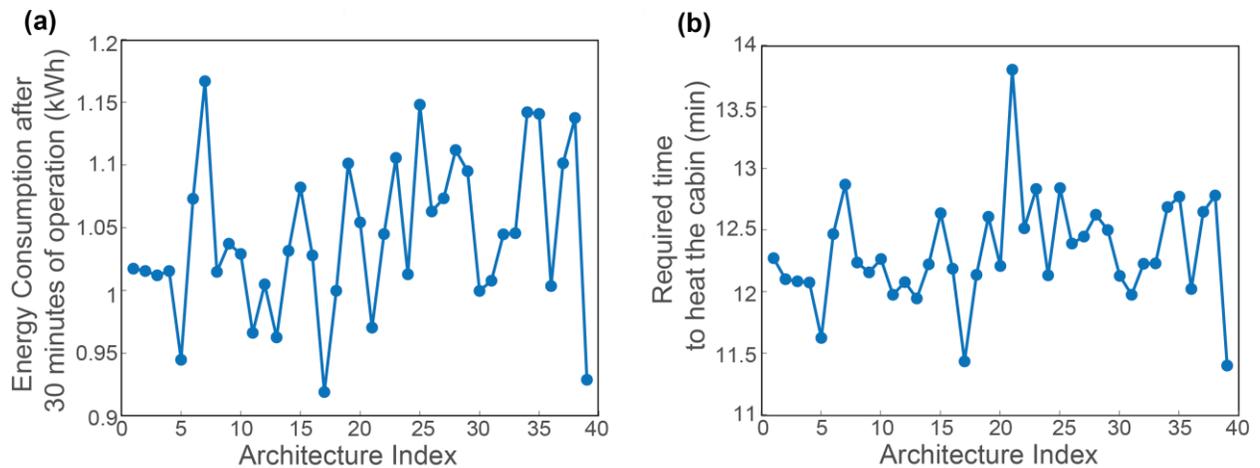

**Figure 13:** The dynamic operating mode test results. **a)** Total energy consumption **b)** The time required to drive the cabin temperature to the comfort temperature

The goal of this study is to minimize both objectives. Figure 14 illustrates the trade-offs between complexity and performance objectives. In this discrete multi-objective optimization problem, the Pareto front and optimal architectures are highlighted in green. As shown, architecture 17 achieves the best performance, while architecture 38 has the lowest complexity factor. Notably, although architecture 39 exhibited the best performance regarding cabin heating time and closely matches architecture 17 in the performance metric, it is excluded from the Pareto front due to its higher complexity.

This analysis demonstrates how the Pareto front and multi-objective framework can guide architecture selection based on the trade-off between complexity and performance. Users can select an architecture depending on their willingness to accept additional complexity to achieve improved performance.

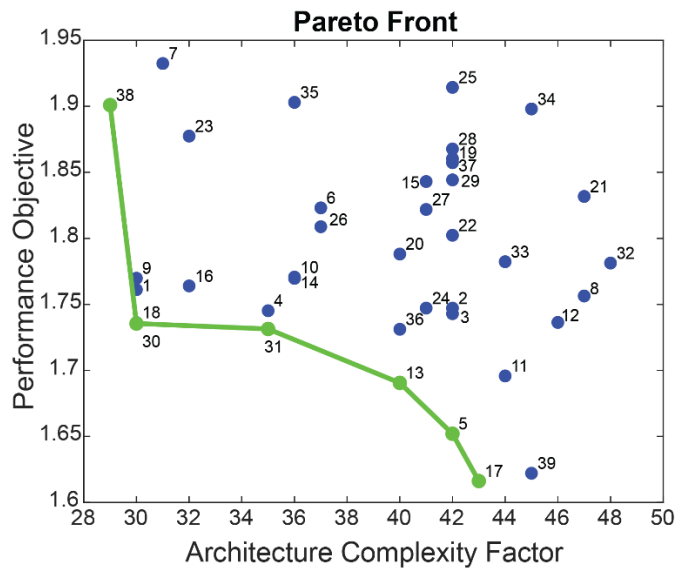

**Figure 14:** Pareto front showing the trade-off between architectural complexity and performance objectives. The horizontal axis represents the complexity factor, capturing the number of components such as valves and junctions, while the vertical axis represents the performance objective, reflecting energy efficiency and cabin heating time. The green points indicate architectures achieving optimal trade-offs.

## 5 CONCLUSIONS

This study developed a framework to facilitate holistic investigation of multi-modal thermal management system architecture for battery electric vehicles. Multi-modal design criteria for battery electric vehicle thermal management system architectures were formulated, and a novel method for automated enumeration of reconfigurable architectures was introduced. Using this approach, 39 unique architectures were generated and modeled automatically in MATLAB Simscape. Multiple test cases and metrics were defined to evaluate these architectures, enabling a detailed performance assessment. A Pareto front analysis of complexity and performance objectives provided a comprehensive comparison of the design candidates. The proposed method serves as a decision support tool for early-stage design, enabling efficient exploration of the architecture design space. Additionally, the automated architecture enumeration tool has the potential to support the design of multi-modal reconfigurable systems in other domains. This study focused on architectures utilizing in-line valves. However, replacing multiple in-line valves connected at the same junction with a single multi-way valve could improve system designs by enhancing practicality and reducing complexity. Future work should integrate multi-way valves directly into the initial enumeration framework, potentially decreasing the overall valve count and further simplifying generated architectures. Additionally, the current analysis employed a limited set of test cases and metrics. Developing a graphical user interface would facilitate broader exploration of the design space by enabling users to define diverse test scenarios and objectives, thus allowing a detailed investigation into how parameters such as vehicle type influence performance and feasibility of enumerated thermal architectures. Furthermore, although a comprehensive analysis was conducted to identify optimal architectures across multiple metrics, the investigation into the underlying reasons behind differing performance among architectures was constrained by a manual process with limited scope. Future research should incorporate advanced data science techniques, such as clustering, feature importance analysis, and explainable AI methods, to systematically uncover and understand the key factors driving these performance differences. This enhanced analytical approach would provide deeper insights, enabling more informed and effective optimization of architectural designs.

# 6 ACKNOWLEDGEMENTS

This work was supported by the National Science Foundation Engineering Research Center for Power Optimization of Electro-Thermal Systems (POETS) through cooperative agreement EEC-1449548. AI-based tools, including ChatGPT, were utilized solely for grammar and language improvements during the preparation of this article. The authors reviewed and edited the manuscript thoroughly after employing these tools and accept full responsibility for its content. N.M. gratefully acknowledges funding support from the International Institute for Carbon Neutral Energy Research (WPI-I2CNER), sponsored by the Japanese Ministry of Education, Culture, Sports, Science, and Technology. Scanning electron microscopy was carried out in part in the Materials Research Laboratory Central Facilities, University of Illinois, Urbana-Champaign.

24. Min, H. *et al.* A thermal management system control strategy for electric vehicles under low-temperature driving conditions considering battery lifetime. *Appl Therm Eng* **181**, 115944 (2020).
25. Ma, J., Sun, Y., Zhang, S., Li, J. & Li, S. Experimental study on the performance of vehicle integrated thermal management system for pure electric vehicles. *Energy Convers Manag* **253**, 115183 (2022).
26. Singh, S., Jennings, M., Katragadda, S., Che, J. & Miljkovic, N. System design and analysis methods for optimal electric vehicle thermal management. *Appl Therm Eng* **232**, 120990 (2023).
27. Yokoyama, A., Osaka, T., Imanishi, Y. & sekiya, S. Thermal Management System for Electric Vehicles. *SAE International Journal of Materials and Manufacturing* **4**, 2011-01–1336 (2011).
28. Vincent George Johnston. EV Multi-Mode Thermal Management System. (2014).
29. Nicholas MANCINI *et al.* Optimal source electric vehicle heat pump with extreme temperature heating capability and efficient thermal preconditioning. (2019).
30. Man Ju Oh, Jae Woong Kim & Sang Shin Lee. HVAC system of vehicle with battery heating and cooling. (2020).
31. Sahni, S. & Bhatt, A. The complexity of design automation problems. in *Proceedings of the seventeenth design automation conference on Design automation - DAC '80* 402–411 (ACM Press, New York, New York, USA, 1980). doi:10.1145/800139.804562.
32. Abtahi, M., Rabbani, M. & Nazari, S. An Automatic Tuning MPC with Application to Ecological Cruise Control. *IFAC-PapersOnLine* **56**, 265–270 (2023).
33. Tian, B. *et al.* WeldMon: A Cost-effective Ultrasonic Welding Machine Condition Monitoring System. in *2023 IEEE 14th Annual Ubiquitous Computing, Electronics & Mobile Communication Conference (UEMCON)* 0310–0319 (IEEE, 2023). doi:10.1109/UEMCON59035.2023.10315983.
34. Eslaminia, A., Meng, Y., Nahrstedt, K. & Shao, C. Federated domain generalization for condition monitoring in ultrasonic metal welding. *J Manuf Syst* **77**, 1–12 (2024).
35. Ruddigkeit, L., van Deursen, R., Blum, L. C. & Reymond, J.-L. Enumeration of 166 Billion Organic Small Molecules in the Chemical Universe Database GDB-17. *J Chem Inf Model* **52**, 2864–2875 (2012).
36. Cilardo, A. & Fusella, E. Design automation for application-specific on-chip interconnects: A survey. *Integration* **52**, 102–121 (2016).
37. Hornby, G. S., Lipson, H. & Pollack, J. B. Generative representations for the automated design of modular physical robots. *IEEE Transactions on Robotics and Automation* **19**, 703–719 (2003).
38. Bayrak, A. E., Ren, Y. & Papalambros, P. Y. Topology Generation for Hybrid Electric Vehicle Architecture Design. *Journal of Mechanical Design* **138**, (2016).
39. Bayat, S. *et al.* Multi-Split Configuration Design for Fluid-Based Thermal Management Systems. *Journal of Mechanical Design* **147**, (2025).
40. Bayat, S. *et al.* Extracting Design Information From Optimized Designs of Power Flow Systems: Application to Multisplit Thermal Management System Configuration. *Journal of Mechanical Design* **147**, (2025).
41. Bayat, S. *et al.* Advancing Fluid-Based Thermal Management Systems Design: Leveraging Graph Neural Networks for Graph Regression and Efficient Enumeration Reduction. (2023).
42. Peddada, S. R. T., Herber, D. R., Pangborn, H. C., Alleyne, A. G. & Allison, J. T. Optimal Flow Control and Single Split Architecture Exploration for Fluid-Based Thermal Management. *Journal of Mechanical Design* **141**, (2019).
43. Buettner, R., Herber, D. R., Abolmoali, P. C. & Patnaik, S. S. An Automated Design Tool for Generation and Selection of Optimal Aircraft Thermal Management System Architectures. in *AIAA Propulsion and Energy 2021 Forum* (American Institute of Aeronautics and Astronautics, Reston, Virginia, 2021). doi:10.2514/6.2021-3718.
44. Herber, D. R., Guo, T. & Allison, J. T. Enumeration of Architectures with Perfect Matchings. *Journal of Mechanical Design* **139**, (2017).
45. Herber, D. R., Allison, J. T., Buettner, R., Abolmoali, P. & Patnaik, S. S. Architecture generation and performance evaluation of aircraft thermal management systems through graph-based techniques. in *AIAA Scitech 2020 Forum* vol. 1 PartF (American Institute of Aeronautics and Astronautics Inc, AIAA, 2020).
50

# APPENDIX A

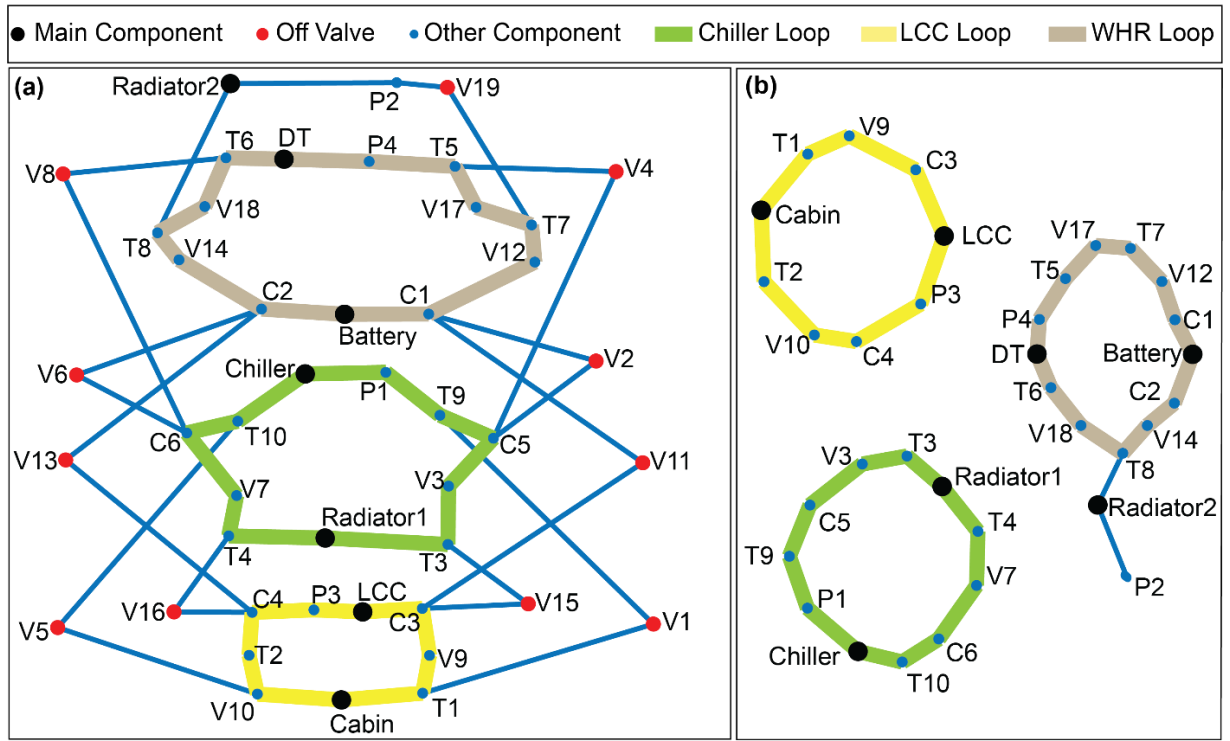

**Figure A. 1:** Layout of Architecture 7. Coolant Loops of mode 2 are highlighted. **a)** Architecture7 fully Pictured. **b)** off valves and coolant lines connected to them were removed from the architecture for a better visualization of the coolant loops.

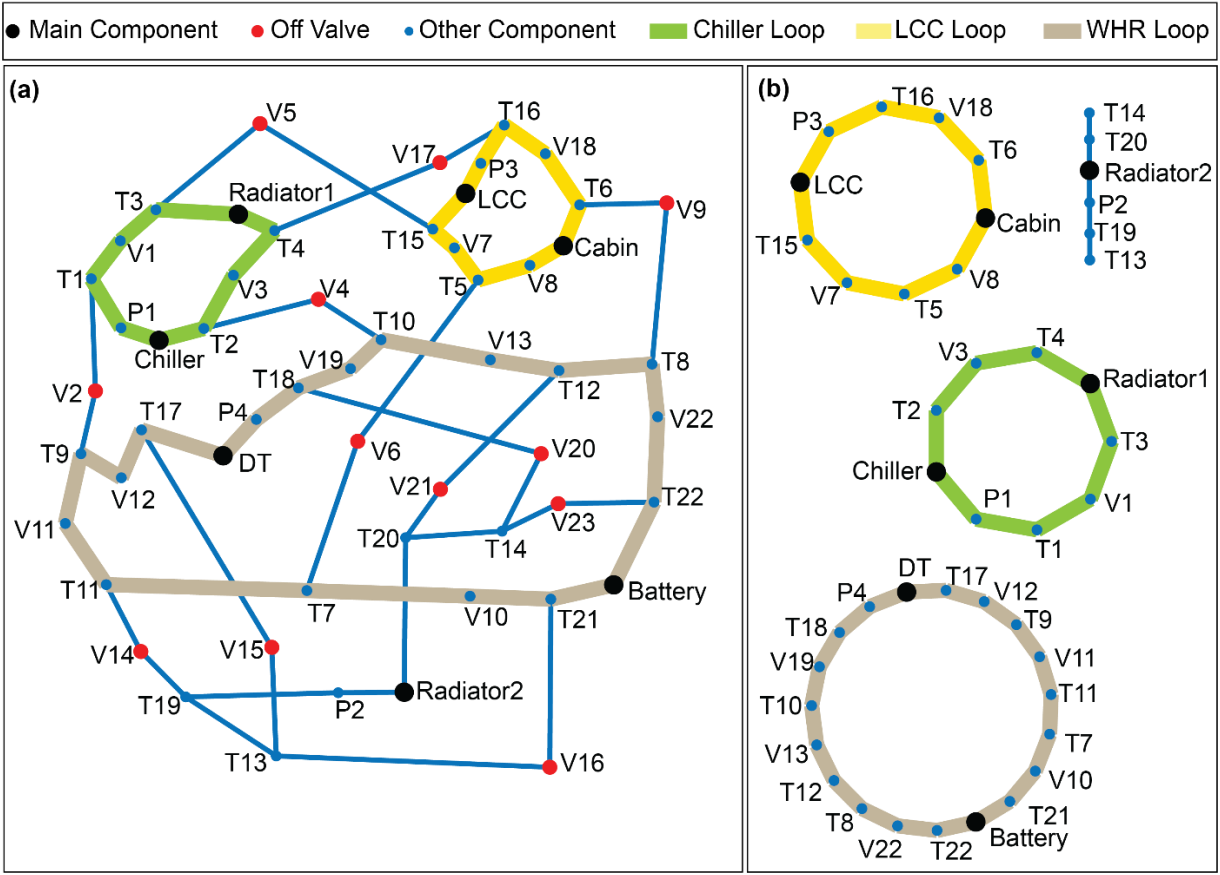

*Figure A. 2*: Layout of Architecture 39. Coolant Loops of mode 2 are highlighted. **a)** Architecture39 fully Pictured. **b)** off valves and coolant lines connected to them were removed from the architecture for a better visualization of the coolant loops.